# Growth from Below: Bilayer Graphene on Copper by Chemical Vapor Deposition

Shu Nie,[1] Wei Wu,[2,3] Shirui Xing,[2,3] Qingkai Yu,[3,4] Jiming Bao,[2] Shin-shem Pei,[2,3,5] and Kevin F. McCarty[1,5]

[1]Sandia National Laboratories, Livermore, CA 94550, USA
[2]Department of Electrical and Computer Engineering, University of Houston, Houston, TX 77204, USA
[3]Center for Advanced Materials, University of Houston, Houston, TX 77204, USA
[4]Ingram School of Engineering and Materials Science, Engineering and Commercialization Program, Texas State University, San Marcos, TX 78666, USA

E-mail: mccarty@sandia.gov, spei@uh.edu

**Abstract.** We evaluate how a second graphene layer forms and grows on Cu foils during chemical vapor deposition (CVD). Low-energy electron diffraction and microscopy is used to reveal that the second layer nucleates and grows next to the substrate, i.e., under a graphene layer. This underlayer mechanism can facilitate the synthesis of uniform *single-layer* films but presents challenges for growing uniform *bilayer* films by CVD. We also show that the buried and overlying layers have the same edge termination.

---

[5] Authors to whom correspondence should be addressed.

## 1. Introduction

In this paper we examine the growth mechanism of graphene bilayers on copper foils during chemical vapor deposition (CVD). Growth by CVD on metals has emerged as a promising technology for graphene synthesis. Single-layer films have been deposited over large areas Cu substrates [1-3]. The CVD growth of bilayer graphene (BLG) at the wafer scale has also been reported [4]. When the two layers are stacked as in graphite, applying an electrical field perpendicular to BLG creates a controllable bandgap [5]. Indeed, Zhang et al. demonstrated that the bandgap can be continuously tuned using BLG field-effect transistors with dual gates [6]. Realizing a bandgap opens the door of digital electronics. The challenge is to develop a BLG technology that gives a uniform and reproducible bandgap over large areas. A better understanding of how BLG grows in CVD will aid this development and also improve the growth of single-layer graphene that is free of BLG.

Films grown by CVD on Cu frequently have isolated grains with discrete regions of single and bilayer graphene. The scanning electron microscopy (SEM) images in figures 1(a) and (b) provide examples. The smaller, darker regions have two graphene layers, the medium-grey regions have one layer and the lightest regions are bare Cu. For such bilayer grains on Cu substrates, the literature [2, 7-10] has often assumed that the smaller layer is on top of the larger layer, like the tiered wedding cake illustrated in figure 2(a). (See, for example, figure 4 of ref. [7] and figure 3 of ref. [10].) If the mechanism of graphene growth is like other materials, this assignment of the layer stacking would be reasonable -- in crystal growth, layers almost always nucleate and grow on top of the crystal [11]. In this "on-top" mechanism, the smaller layer in figure 2(a) is the second layer to form. However, Tontegode and co-workers used Auger electron spectroscopy to suggest that graphene grows with an *inverted* wedding-cake structure, as in figure 2(b). That is, they proposed that the second and subsequent graphene layers grow next to the substrate during segregation from Re and when C is deposited on Ir(111) [12, 13]. Recent microscopy has confirmed this "underlayer" mechanism on Ir(111) [14] and Ru(0001) [14, 15]. The most detailed analysis has been performed on Ir(111), where low-energy electron diffraction (LEED) and low-energy electron microscopy (LEEM) directly showed that a graphene layer nucleated and grew under a single layer. This growth next to the substrate occurred when C was segregating from the Ir substrate or when C was deposited on top of a complete single layer [14].

On the other hand, the community has not accepted the underlayer mechanism for CVD on Cu. Indeed, a recent paper notes that it "is not possible" for Cu(111) [8]. In fact there are reasons that bilayer growth maybe different in CVD, particularly on Cu foils. First, the negligible carbon solubility in Cu prevents graphene growth by segregation [1]. Second, CVD relies on the substrate itself to generate the growth species by catalyzing hydrocarbon decomposition [16]. The growth species must diffuse from the bare metal to the graphene. And growth stops once the surface is covered by graphene. In contrast, during segregation and C deposition the active growth species (C adatoms) is uniformly supplied over the entire surface and growth continues when the surface is covered by graphene. In this report we show that the underlayer mechanism also occurs for CVD on Cu.

The key in distinguishing between the two growth mechanisms is determining whether the smaller layer is on top (figure 2(a)) or next to the substrate (figure 2(b)). This is a challenging analytical problem. For example, configurations 2(a) and 2(b) will image very similarly in scanning probe microscopy. We show how the layers are stacked in BLG grains grown by CVD on Cu foils using diffraction from low-energy electrons, whose small penetration depth allows the top layer to be distinguished from the layer next to the substrate. We find that the smaller graphene layers are buried below the larger layers, i.e., the grains are stacked like an inverted wedding cake (figure 2(b)). This observation offers persuasive evidence that graphene always nucleates and grows next to the Cu, even when it is already covered by graphene. We discuss the implications of underlayer growth for achieving uniform single-layer and BLG. We also show that the edges of the buried and overlying layers have the same termination.

## 2. Results

We deposited BLG and few layer graphene (FLG) on Cu foils (25 µm thick foils, 99.8%, Alfa Aesar) by ambient-pressure CVD [2, 17]. Prior to growth, the Cu foils were annealed in Ar and $H_2$ at 1050 °C for 30 min, which cleans the surface and increases the grain size. Our previous work established that $CH_4$ concentrations lower than 20 ppm with 1.3% $H_2$ at 1050 °C result in single-layer graphene, while higher $CH_4$ concentrations lead to a large density of multi-layer grains [2]. For this study, we used CVD conditions (950 °C with 1.3% $H_2$ in Ar at a total flow rate of 1500 sccm), including a higher $CH_4$ concentration (30 ppm), that were optimized to give discrete grains consisting two or three faceted layers. (Figure 2 in ref. [2] shows SEM images of single-layer graphene produced at lower $CH_4$ concentrations.) The $CH_4$ flow was stopped after 30 min and the sample was then quickly cooled under the protection of the Ar and $H_2$. The foil was then transferred in air to the LEEM, where it was annealed at 200 °C before analysis.

Our method to identify the graphene layer next to the substrate first uses LEEM to distinguish single-layer and bilayer graphene and then employs LEED to analyze bilayer regions whose two layers are misaligned rotationally. In LEEM, an interference phenomenon in electrons reflected from a film on a substrate gives a direct measure of the film thickness in atomic layers [18]. Figure 3 shows how electron reflectivity changes as a function of electron energy. BLG gives a single pronounced minimum while three-layer graphene has two minima.[6] Using the ability to determine layer thickness, we then analyze the stacking of bilayers whose two graphene layers are rotated with respect to each other. The SEM images in figures 1(a) and (b) show two graphene grains whose centers are bilayers. We find that the edges of the two stacked hexagonal layers in a grain can be parallel to each other, figure 1(a), or rotated, figure 1(b), consistent with other reports [9]. The photoelectron emission microscopy (PEEM) image at a lower magnification in figure 1(c) also shows examples where the edges of smaller hexagonal shapes (dark) are rotated with respect to the edges of the larger hexagonal grains (bright). We next show that concentric layers with non-parallel edges are indeed rotated with respect to each other. Then the diffraction spots from each layer are separated and distinguishable, and the layer that is next to the substrate can be determined [14].

Figure 4 provides an example analysis that establishes whether the second layer nucleated above (figure 2(a)) or below the initial layer (figure 2(b)). The dark-grey, distorted hexagon near the center of figure 4(a) is BLG, as established by the electron reflectivity (i.e., figure 3). The surrounding bright hexagonal region is a single layer whose edges are not aligned with those of the smaller, concentric layer. Figure 4(d) shows diffraction from the single-layer region marked by the red box. Red arrows mark the graphene spots.[7] Diffraction from the BLG region in the blue box, figure 4(e), has the same graphene spots as the single layer. But there is an additional set of weak but sharp graphene spots,[8] marked blue. This second set of spots comes from the smaller layer of the BLG. We use diffraction intensities to discriminate whether the grain's smaller layer is above or below its larger layer [i.e., figure 2(a) or (b)]. Diffraction from the smaller layer is clearly much weaker in figure 4(e), a consequence of the strong attenuation of the 50-eV electrons during transmission through a single graphene layer. Thus, the smaller layer is below the larger layer. Figure 2(b) illustrates the stacking. In the example in figure 4, the two sets of 6-fold spots are rotated by ~23°, which is the relative rotation between the lattices of the two layers.

---

[6] The small oscillations in the single-layer reflectivity are not interference features.

Additional observations support the conclusion that the smaller layer of the bilayer region is next to the Cu foil. Figure 4(b) is a dark-field LEEM image from the graphene diffraction spot that is red-circled in figure 4(e). The outer, single-layer region of the hexagonal grain is bright in the image since this region diffracts into the selected spot. The center BLG region is also bright. This is entirely consistent with the schematic in figure 2(b). That is, the sheet of the single-layer region is also the top sheet of the bilayer region. So both regions diffract electron into the "red" spot. The surrounding grains are dark because they are rotated in plane relative to the center grain. Consistently, only the grain's bilayer region has intensity in a dark-field image (figure 4(c)) formed from a buried-layer spot (circled in blue in figure 4(e)). (Weak diffraction from the buried layer gives a dim image.) Another check comes from examining the bottom of figure 4(c). There other single-layer regions are bright because they are closely aligned rotationally (within ~2°) with the center grain's buried layer, as the LEED pattern in the insert shows.[9] The last check is that diffraction from a misaligned buried layer becomes brighter relative to diffraction from the top layer as the penetration depth increases with electron energy.[10]

---

[7] The other spots come from the graphene on the inclined Cu facets (see below) or from the graphene/Cu moiré.
[8] The weak and strong diffraction spots have similar sharpness. Thus, differences in graphene disorder do not cause the intensity change.

The bright/dark stripes in the images in figure 4 result from an array of two facet types on the Cu grain. From the diffraction's energy dependence [19], the angle between the facets is ~13.5°, a value consistent with the array being alternating (100) and (410) facets. Indeed Perdereau and Rhead found that graphene grown by C deposition caused vicinal Cu(100) to facet into orientations including (410) [20].

## 3. Discussion

We used the above methodology to examine seven bilayer regions whose layers were rotationally misaligned. All these regions had the smaller layers of bilayer grains next to the substrate, below larger layers (figure 2(b)). So clearly the smaller layers are growing below the overlying layers. Because of the high gas pressures required for CVD on Cu, in-situ observations of the buried layer nucleating cannot be obtained with LEEM, unlike for Ir and Ru [14]. However, the most straightforward explanation for the inverted wedding-cake structure is that new layers not only grow but also nucleate next to the substrate.[11] Figure 2(c) illustrates the mechanism.

---

[9] The six-fold set of graphene spots in the insert are aligned with the weak, blue-marked spots in figure 4(e), which come from the center grain's buried layer.
[10] For example, at 50 eV, the weak spots were ~9% as intense as the bright spots. At the more penetrating energy of 174 eV, the ratio was ~27%.

Underlayer nucleation also provides a natural explanation for the concentric stacking of the layers seen in figures 1, 3, and 4 and the literature [2, 10]. This concentricity suggests that all layers in a grain nucleate at a common site distinguished by an impurity or special topography. In the underlayer mechanism, new layers nucleate next to the substrate, so the site remains active even when covered by graphene.[12]

Consideration of the underlayer mechanism also aids understanding the processes that are being used to make uniform single layers. If the hydrocarbon concentration in the gas feed is low initially, only single layers form. For fixed process conditions, the graphene coverage increases at a rate that is proportional to the coverage of bare metal [16]. Then the layer never actually completes – the coverage approaches saturation exponentially in time. So how can the process be speed up without forming multilayers? Nucleating a buried layer requires C adatoms to diffuse from the bare Cu (see figure 2(c)) to the center of the single-layer grain.[13] Building up the C concentration sufficiently to nucleate the buried layer is easier when this diffusion length is short. This occurs early in CVD, when the supersaturation is the highest [22] and the edge of the single layer is closer to the nucleation site. (That is, while the length of graphene edges, which consume the growth species, is small.) After the grains have expanded, though, the hydrocarbon concentration can be increased without forming additional layers. If this higher concentration was used initially, BLG graphene regions would form early in growth, as in the sample we analyzed. Wu et al. have used this scheme to speed the completion of a uniform single layer [2].

---

[11] In the alternative scenario, the new layer nucleating on top (figure 2(a)), the edge of the top layer would be further away from the source of the growth species, the bare Cu [1, 16], than the edge of the larger bottom layer. (The latter faces the exposed Cu). Since, it is unlikely that the top layer can grow quickly enough to bury the bottom layer, this mechanism seems implausible. Instead, all observations are explained by the new layer growing and also nucleating next to the substrate.

[12] Also, the only significant difference between the nucleating on the bare Cu and below a graphene sheet is that latter requires debonding graphene from the substrate. Since Cu binds graphene weakly [21], this debonding does not have a high energetic requirement and the C adatom concentrations needed for nucleating a single and subsequent layers do not differ greatly.

[13] The fact that C atoms diffuse past the free graphene edge suggests that there is an energy barrier for attaching C to the edge, as is known for Ru(0001) and Ir(111) substrates [16, 22].

We next discuss the detrimental consequences of the underlayer mechanism that are specific to CVD growth. During C segregation from the substrate or C deposition, new layers can form even after the surface is covered by graphene. In contrast, in CVD, the growth species is generated only where there is bare Cu to decompose the hydrocarbon feed gas. Thus, graphene growth stops when there is no bare Cu [1, 16].[14] Then the buried layer has to be grown before the single layer completes. For our CVD conditions, however, the buried layer appears to grow significantly slower, as evidenced by the small fraction that is bilayer (see figures 1 and 4). Overall a simple CVD process faces two serious challenges to synthesizing a uniform bilayer -- nucleating the buried layer soon after the single layer and growing both sheets at essentially the same velocity.

While measuring statistics was not a focus of this study, we note that rotational misalignment in the top and buried layers was not rare in sample we analyzed (see figure 1(c)). The relative rotations of the seven misaligned bilayers we analyzed were -5, 16, 23, 25, 28, 29, and 30°, with an estimated error of ±2°. Such misaligned bilayers cannot have the AB (Bernal) stacking of layers present in graphite. The observations that misalignment occurs and the spread of relative rotations is wide suggest that neither the substrate nor the overlying graphene strongly locks the buried layer into the same orientation as the overlying layer.

---

[14] For example, Wu et al. found that increasing the methane concentration by tenfold after completing a single layer did not cause new-layer formation [2].

Finally, the facets of the buried and overlying layers are rotated by the same amount as the lattices, within measurement error. In figure 4, for example, the LEED spots of the two layers are rotated by 23°. And the two sets of facets in the images are rotated by the same value. Thus, the facets of both layers lie along the same crystallographic direction of graphene.[15] Then the crystallographic alignment in bilayer stacks can be simply measured from images like figure 1. Furthermore, ignoring possible reconstructions [23], the buried and top layers have the same edge termination. For hexagonal grains of single-layer graphene grown by CVD on Cu, this termination is predominantly of zigzag type [24].

## 4. Summary

The underlayer mechanism of figure 2(b) occurs on a variety of materials and independent of whether the C comes from segregation (Re [12, 13], Ru [14, 15], Ir [14] and Ni [25]), CVD (Cu), carbon deposition (Ir [13, 14]) and even the decomposition of the substrate, as for SiC [26, 27]. This common behavior has a simple origin – except for being incorporated into graphene, the most stable binding site for the carbon atoms needed for growth is at the substrate, not on the graphene [22, 28]. In segregation and decomposition, the growth species is generated at the substrate/film interface and accumulates there. In CVD and C deposition, the carbon diffuses to the substrate/film interface, leading to the nucleation and growth of buried layers (figure 2(c)). Recognition of the underlayer mechanism aids understanding how uniform single layers can be synthesized by controlling the initial supersaturation [2]. In contrast, growth from below seriously constrains the ability to make continuous BLG by simple CVD, where growth stops once the overlying layer is complete. So the two layers must be completed at the same time, which is a challenge since the edge of the buried layer most certainly advances more slowly than the overlying layer's exposed edge. However, some applications do not require continuous films of BLG. Instead single-crystal BLG is only needed in the local regions where the active devices reside. This could be achieved using the approach of Yu et al. [24, 29], where graphene grains grow at the predetermined locations of carbon-rich seeds fabricated on the Cu substrate prior to CVD.

---

[15] Consistently, when diffraction established that both layers are crystallographically aligned (i.e., only one set of graphene diffraction spots), both facet sets were also aligned.

Acknowledgement: The work at Sandia National Laboratories was supported by the Office of Basic Energy Sciences, Division of Materials Sciences and Engineering of the US DOE under Contract No. DE-AC04-94AL85000. S.S.P., J.M.B., W.W., and S.R.X. acknowledge support from the Delta Electronics Foundation and UH CAM. Q.K.Y. acknowledges support from NSF under Grant No. DMR-0907336. The authors thank N. C. Bartelt for informative discussions.

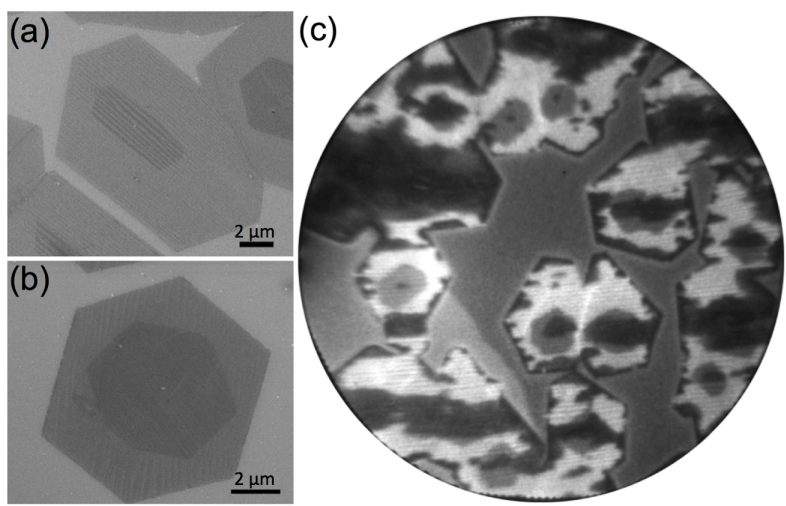

Figure 1. (a,b) SEM and (c) PEEM images of graphene on Cu. In the SEM images, the medium-grey regions are faceted single-layer sheets. The darker regions, which are also faceted, have two graphene layers. PEEM field of view is 50 µm.

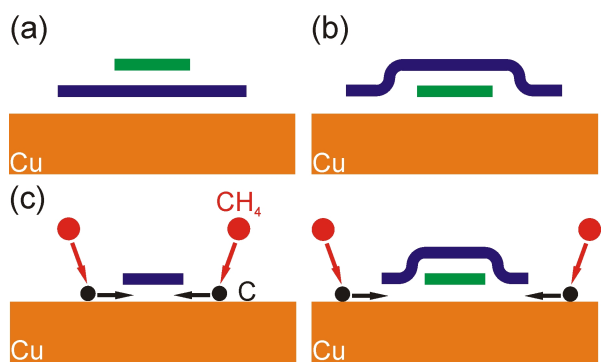

Figure 2. Cross-sectional schematics of graphene layers on a Cu substrate. (a) On-top growth. Graphene layers stacked like a tiered wedding cake, where the smaller sheet of the bilayer is on top of the larger sheet. This arrangement is expected for typical crystal growth. (b). Underlayer growth. Graphene layers stacked like an inverted wedding cake, where the smaller layer is under the larger layer. (c) Underlayer mechanism of nucleation and growth during CVD on Cu. The new graphene layer (green) nucleates below the single layer (blue), giving an inverted wedding cake. Methane decomposes on the bare Cu, generating C adatoms that diffuse under the graphene sheet.

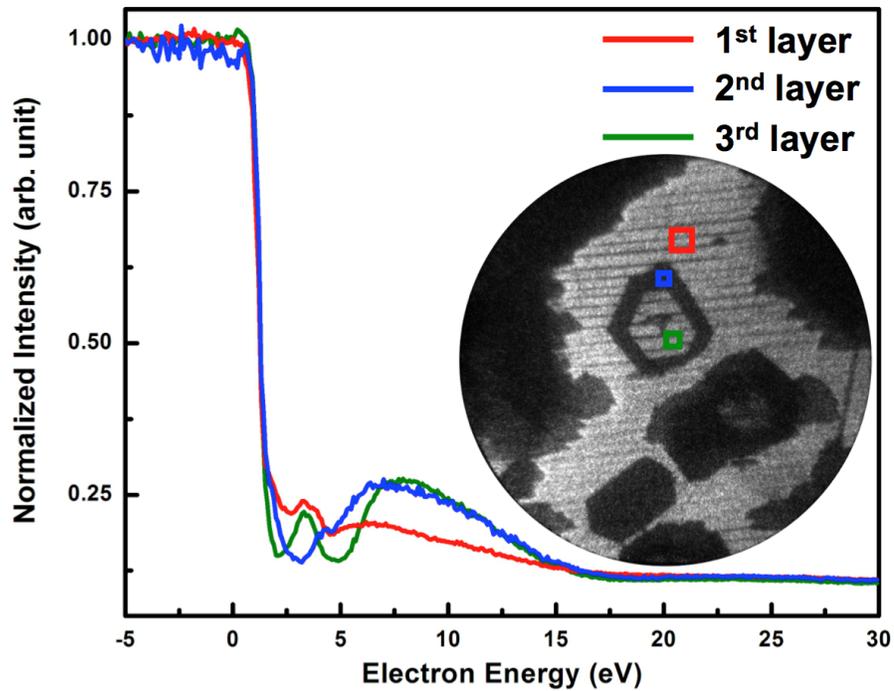

Figure 3. Electron reflectivity vs. electron energy for one - three graphene layers on Cu foil. Measurement regions are color coded in the LEEM image (field of view is 15 μm). One, two, and three layer have no pronounced minimum, a single minimum at ~3 eV, or two minima, respectively.

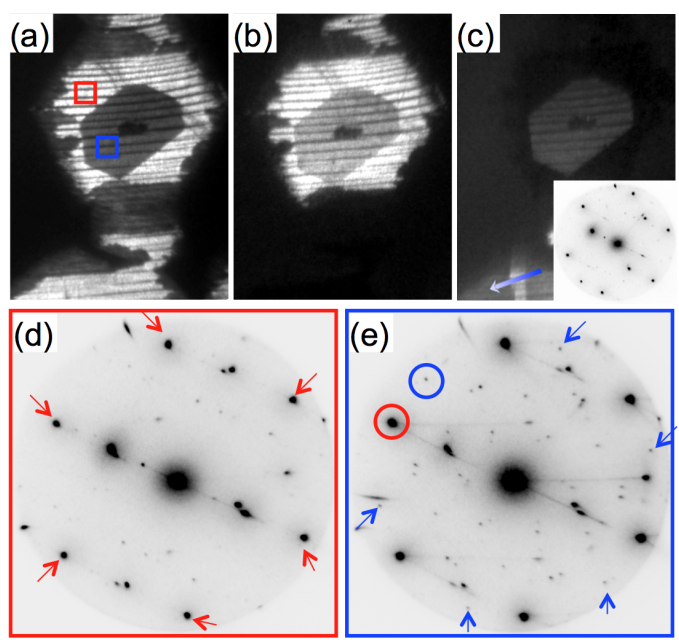

Figure 4. (a) Bright-field and (b-c) dark-field LEEM images (11.5 x 15 μm). The grey distorted hexagon near the center of (a) is BLG and the surrounding bright region is single layer. The facets of the grey and bright hexagons are rotated. (d) Selected-area LEED from the single-layer region marked by the red box in (a). Red arrows mark the single set of 6-fold graphene spots. (e) Diffraction from the BLG region marked by the blue box. The weak graphene spots marked in blue come from the buried layer. Dark-field image (b) is from the top-layer diffraction spot circled in red in (e). Dark-field image (c) is from the buried-layer diffraction spot circled in blue. The insert diffraction pattern in (c) is from the single-layer region marked by the arrow. Images a-c and diffraction patterns obtained using 3.9, 67, 65 and 50 eV electrons, respectively.